\documentclass[reprint,superscriptaddress,nofootinbib,amsmath,amssymb,aps,prd,]{revtex4-2}

\usepackage{graphicx}
\usepackage{dcolumn}
\usepackage{bm}
\usepackage{hyperref}
\usepackage{color}
\usepackage{soul}

\graphicspath{{Figures/}}

\newcommand{\LCDM}{{$\rm{\Lambda CDM}$}}

\newcommand{\planck}{\textit{Planck}}

\newcommand{\BK}[1]{{BK#1}}
\newcommand{\lowlEB}{\texttt{lowlEB}}

\newcommand{\commander}{{\sc commander}}
\newcommand{\hillipop}{\texttt{HiLLiPoP}}
\newcommand{\lowl}{\mbox{low-$\ell$}}

\providecommand{\sorthelp}[1]{}
\newcommand{\aap}{Astron. astrophys.}
\newcommand{\mnras}{Mon. Not. R. Astron. Soc.}
\newcommand{\jcap}{J. Cosmol. Astropart. Phys.}
\newcommand{\araa}{Ann. Rev. Astron. Astrophys.}

\begin{document}

\title{Improved limits on the tensor-to-scalar ratio using BICEP and \textit{Planck}}

\author{M.~Tristram}
\affiliation{Universit\'{e} Paris-Saclay, CNRS/IN2P3, IJCLab, 91405 Orsay, France}

\author{A.~J.~Banday}
\affiliation{IRAP, Universit\'{e} de Toulouse, CNRS, CNES, UPS, (Toulouse), France}

\author{K.~M.~G\'{o}rski}
\affiliation{Jet Propulsion Laboratory, California Institute of Technology, 4800 Oak Grove Drive, Pasadena, California, U.S.A.}
\affiliation{Warsaw University Observatory, Aleje Ujazdowskie 4, 00-478 Warszawa, Poland}

\author{R.~Keskitalo}
\affiliation{Computational Cosmology Center, Lawrence Berkeley National Laboratory, Berkeley, California, U.S.A.}
\affiliation{Department of Physics and Space Sciences Laboratory, University of California, Berkeley, California, U.S.A.}

\author{C.~R.~Lawrence}
\affiliation{Jet Propulsion Laboratory, California Institute of Technology, 4800 Oak Grove Drive, Pasadena, California, U.S.A.}

\author{K.~J.~Andersen}
\affiliation{Institute of Theoretical Astrophysics, University of Oslo, Blindern, Oslo, Norway}

\author{R.~B.~Barreiro}
\affiliation{Instituto de F\'{\i}sica de Cantabria (CSIC-Universidad de Cantabria), Avda. de los Castros s/n, Santander, Spain}

\author{J.~Borrill}
\affiliation{Computational Cosmology Center, Lawrence Berkeley National Laboratory, Berkeley, California, U.S.A.}
\affiliation{Space Sciences Laboratory, University of California, Berkeley, California, U.S.A.}

\author{L.~P.~L.~Colombo}
\affiliation{Dipartimento di Fisica, Universit{\`a} degli Studi di Milano, Via Celoria, 16, Milano, Italy}

\author{H.~K.~Eriksen}
\affiliation{Institute of Theoretical Astrophysics, University of Oslo, Blindern, Oslo, Norway}

\author{R.~Fernandez-Cobos}
\affiliation{Dpto. de Matem\`{a}ticas, Estad\'{i}stica y Computaci\'{o}n, Universidad de Cantabria, Avda. de los Castros s/n, E-39005 Santander, Spain.}

\author{T.~S.~Kisner}
\affiliation{Computational Cosmology Center, Lawrence Berkeley National Laboratory, Berkeley, California, U.S.A.}
\affiliation{Department of Physics and Space Sciences Laboratory, University of California, Berkeley, California, U.S.A.}

\author{E.~Mart\'{\i}nez-Gonz\'{a}lez}
\affiliation{Instituto de F\'{\i}sica de Cantabria (CSIC-Universidad de Cantabria), Avda. de los Castros s/n, Santander, Spain}

\author{B.~Partridge}
\affiliation{Haverford College Astronomy Department, 370 Lancaster Avenue, Haverford, Pennsylvania, U.S.A.}

\author{D.~Scott}
\affiliation{Department of Physics \& Astronomy, University of British Columbia, 6224 Agricultural Road, Vancouver, British Columbia, Canada}

\author{T.~L.~Svalheim}
\affiliation{Institute of Theoretical Astrophysics, University of Oslo, Blindern, Oslo, Norway}

\author{I.~K.~Wehus}
\affiliation{Institute of Theoretical Astrophysics, University of Oslo, Blindern, Oslo, Norway}

\date{\today} 

\begin{abstract}
We present constraints on the tensor-to-scalar ratio $r$ using a combination of BICEP/Keck 2018 (BK18) and \planck\ PR4 data allowing us to fit for $r$ consistently with the six parameters of the \LCDM\ model. We discuss the sensitivity of constraints on $r$ to uncertainties in the \LCDM\ parameters as defined by the \planck\ data. In particular, we are able to derive a constraint on the reionization optical depth $\tau$ and thus propagate its uncertainty into the posterior distribution for $r$. While \planck\ sensitivity to $r$ is slightly lower than the current  ground-based measurements, the combination of \planck\ with \BK{18} and baryon-acoustic-oscillation data yields results consistent with $r=0$ and tightens the constraint to $r < 0.032$ at 95\% confidence.
\end{abstract}

\maketitle

\section{\label{sec:intro}Introduction}
Introduced in order to resolve problems within the Big-Bang cosmological model (such as the horizon, flatness, and magnetic-monopole problems), inflation also naturally provides the seeds for generating primordial matter fluctuations from quantum fluctuations (see for instance Ref.~\cite{kamionkowski16} and references herein).

Measurements of the cosmic microwave background (CMB) allow constraints to be placed on the amplitude of the tensor perturbations that are predicted to be generated by primordial gravitational waves during the inflationary epoch, leaving some imprints on the CMB anisotropies \cite{Seljak97a,Kamionkowski97,Seljak97b,HuWhite1997}.
Over the last decade, while no primordial signals have been discovered, significant improvements on the upper limit for the tensor-to-scalar ratio $r$ have progressively led to the constraint becoming lower than a few percent in amplitude: $r<0.11$ in 2013 using only temperature data from \planck~\cite{planck2013-p11}; $r<0.12$ in 2015 using polarization from BICEP/Keck and \planck~\cite{BKP} to debias the initially claimed detection from BICEP/Keck in 2014, $r=0.2^{+0.07}_{-0.05}$~\cite{BicepDetection}; $r<0.09$ in 2016 using BICEP/Keck and \planck~\cite{planck2014-a15}; $r<0.07$ in 2018 using BICEP/Keck 2015 data~(\BK{15},~\cite{BK15}); $r<0.065$ in 2019 using \planck\ in combination with BK15~\cite{planck2016-l06}; $r<0.044$ in 2021 using \planck\ in combination with BK15~\cite{planck_tensor}; and $r<0.036$ in 2021 using the latest BICEP/Keck data (\BK{18},~\cite{BK18}).
\begin{figure}[htpb!]
\centering
\includegraphics[width=8.5cm,height=6.8cm]{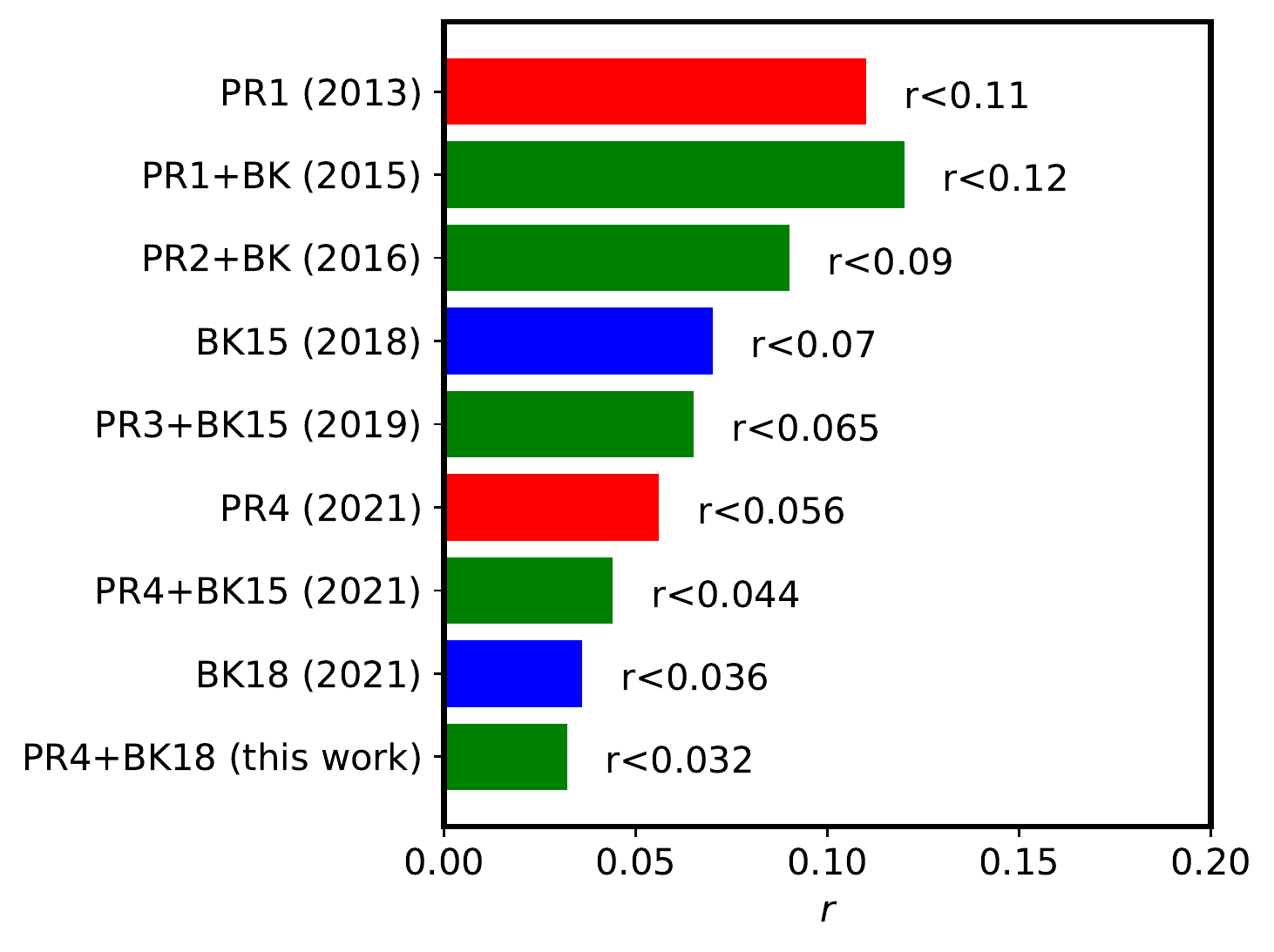}
\caption{History of constraints on the tensor-to-scalar ratio~$r$ (\planck\ PR1~\cite{planck2013-p11},  \planck\ PR1+BK~\cite{BKP}, \planck\ PR2+BK~\cite{planck2014-a15}, BK15~\cite{BK15}, \planck\ PR3+BK15~\cite{planck2016-l06}, \planck\ PR4~\cite{planck_tensor}, \planck\ PR4+BK15~\cite{planck_tensor}, BK18~\cite{BK18}, \planck\ PR4+BK18 this work). Upper limits are given at 95\,\% CL.}
\label{fig:history}
\end{figure}

In this paper, we first discuss the impact of uncertainties in \LCDM\ parameters for constraints on $r$ derived from the latest BICEP/Keck data (\BK{18}) \cite{BK18} alone. Then we add in data from the latest \planck\ release (PR4) \cite{planck2020-LVII} in order to provide the best currently available constraint on the tensor-to-scalar ratio $r$.

\section{\label{sec:model}Cosmological model}
The cosmological model used in this paper is based on adiabatic, nearly scale-invariant perturbations. It has been established as the simplest model that is consistent with the different cosmological probes and in particular with the CMB \cite{planck2016-l06}.

The standard \LCDM+r model includes 6+1 parameters. 
Power spectra for scalar and tensor modes are parameterized by power laws with no running and so the parameters include the scalar amplitude $A_{\rm s}$ and the spectral scalar index $n_{\rm s}$, while the spectral index for the tensor mode $n_{\rm t}$ is set using single-field slow-roll inflation consistency. The amplitudes and the tensor-to-scalar power ratio, $r \equiv A_{\rm t}/A_{\rm s}$, are evaluated at a pivot scale of 0.05\,Mpc$^{-1}$.
Three other parameters ($\Omega_{\rm b}h^2$, $\Omega_{\rm c}h^2$, and $\theta_{\ast}$) determine the linear evolution of perturbations after they re-enter the Hubble scale.
Finally, the reionization is modeled with a widely-used step-like transition between an essentially vanishing ionized fraction at early times, to a value of unity at low redshifts. The transition is modeled using a $\tanh$ function with a non-zero width fixed to $\Delta z = 0.5$~\cite{lewis08}. The reionization optical depth $\tau$ is then directly related to the redshift at which this transition occurs.

The CMB power spectra are generated using the Boltzmann-solver code \textsc{camb}~\cite{Lewis:1999bs,Howlett:2012mh}. 
We sample the likelihood combinations using the \textsc{cobaya} framework \cite{cobaya} with fast and efficient Markov chain Monte Carlo sampling methods described in Refs.~\cite{Lewis:2002ah} and \cite{Lewis:2013hha}. All the likelihoods that we use are publicly available on the \textsc{cobaya} web site\footnote{\href{https://cobaya.readthedocs.io}{cobaya.readthedocs.io}} and are briefly described in the next section.

\section{Planck likelihoods}
We use the polarized likelihood at large scales, \lowlEB, described in Ref.~\cite{planck_tensor} and available on github.\footnote{\href{https://github.com/planck-npipe}{github.com/planck-npipe}} Specifically, it is a \planck\ \lowl\ polarization likelihood based on cross-spectra using the Hamimeche-Lewis approximation~\cite{hamimeche08,mangilli15}.
Using this formalism, the likelihood function consistently takes into account the two polarization fields $E$ and $B$ (including $EE$, $BB$, and $EB$ power-spectra), as well as {\it all\/} correlations between multipoles and modes. It is important to appreciate that such correlations are relevant at large angular scales where cut-sky effects and systematic residuals (both from the instrument and from the foregrounds) are important.
The cross-spectra are calculated on component-separated CMB ``detset'' maps processed by \commander\ from the \planck\ PR4 frequency maps, on 50\,\% of the sky. The sky fraction is optimized in order to obtain maximum sensitivity (and lowest sample variance), while ensuring low contamination from residual foregrounds.
The covariance matrix is estimated from the PR4 Monte Carlos. The statistical distribution of the recovered $C_\ell$s naturally includes the effect of all components included in the Monte Carlo, namely the CMB signal, instrumental noise, \planck\ systematic effects incorporated in the PR4 simulations (see Ref.~\cite{planck2020-LVII}), component-separation uncertainties, and foreground residuals. 
In the case of \planck, we are not able to analytically predict the shape of the full covariance matrix for component-separated maps. However, analytical predictions exist for the covariance of instrumental noise in low-resolution individual-frequency maps. Analysis of these matrices highlights non-trivial structures in the harmonic space noise, whose covariance cannot be approximated as diagonal. Since component-separated maps are a combination of the input frequency maps, part of these structures will carry over into the final covariance matrix, on top of any additional correlations induced by systematic effects, masking, and foreground residuals that cannot be modeled analytically but only reconstructed via simulations. Given these considerations, we cannot apply any type of simplifying ``conditioning'' (such as setting off-diagonal elements to zero), as done for some ground-based experiments, nor do we wish to make such assumptions about the data.
In this paper, unlike previous CMB work to our knowledge, we now marginalize the likelihood over the unknown true covariance matrix (as proposed in Ref.~\cite{sellentin16}) in order to propagate the uncertainty in the estimation of the covariance matrix caused by a limited number of simulations. This provides us with a likelihood that is unbiased and accurate for the estimation of the uncertainty. The robustness of the results is discussed further in the Appendix.

At large angular scales in temperature, we make use of the \planck\ public \lowl\ temperature-only likelihood, based on the CMB map recovered from the component-separation procedure (specifically \commander) described in detail in Ref.~\cite{planck2016-l05}.

At small scales, we use the \planck\ \hillipop\ likelihood, which can include the $TT$, $TE$, and/or $EE$ power spectra computed on the PR4 detset maps at 100, 143, and 217\,GHz. The likelihood is a spectrum-based Gaussian approximation, with semi-analytic estimates of the $C_\ell$ covariance matrix based on the data. The model consists of a linear combination of the CMB power spectrum and several foreground residuals, including Galactic dust, cosmic infrared background, thermal and kinetic Sunyaev-Zeldovich (SZ) effects, SZ-CIB correlations, and unresolved point sources. For details, see Refs.~\cite{planck_tensor} and \cite{planck2013-p08,planck2014-a13,couchot2017}.

\section{BICEP/Keck likelihood}
We use the publicly available BICEP/Keck likelihood (\BK{18}) corresponding to the data taken by the BICEP2, Keck Array, and BICEP3 CMB polarization experiments, up to and including the 2018 observing season~\cite{BK18}.
The format of the likelihood is identical to the one introduced in Refs.~\cite{BKP} and \cite{BK15};
it is a Hamimeche-Lewis approximation \cite{hamimeche08} to the joint likelihood of the ensemble of $BB$ auto- and cross-spectra taken between the BICEP/Keck (two at 95, one each at 150 and 220\,GHz), \textit{WMAP} (23 and 33\,GHz), and \planck\ (PR4 at 30, 44, 143, 217, and 353\,GHz) frequency maps.
The effective coverage is approximately $400\,{\rm deg}^2$ (which corresponds to 1\,\% of the sky) centered on a region with low foreground emission.
The data model includes Galactic dust and synchrotron emission, as well as correlations between dust and synchrotron.
\\

In the following, we neglect correlations between the BICEP/Keck and \planck\ data sets. This is justified because the BK18 spectra are estimated on 1\,\% of the sky, while the \planck\ analysis is derived from 50\,\% of the sky.

\section{Impact of $\mathbf{\Lambda}$\bf{CDM} uncertainty} 
\begin{figure}[htbp!]
\centering
\includegraphics[width=8.6cm,height=6cm]{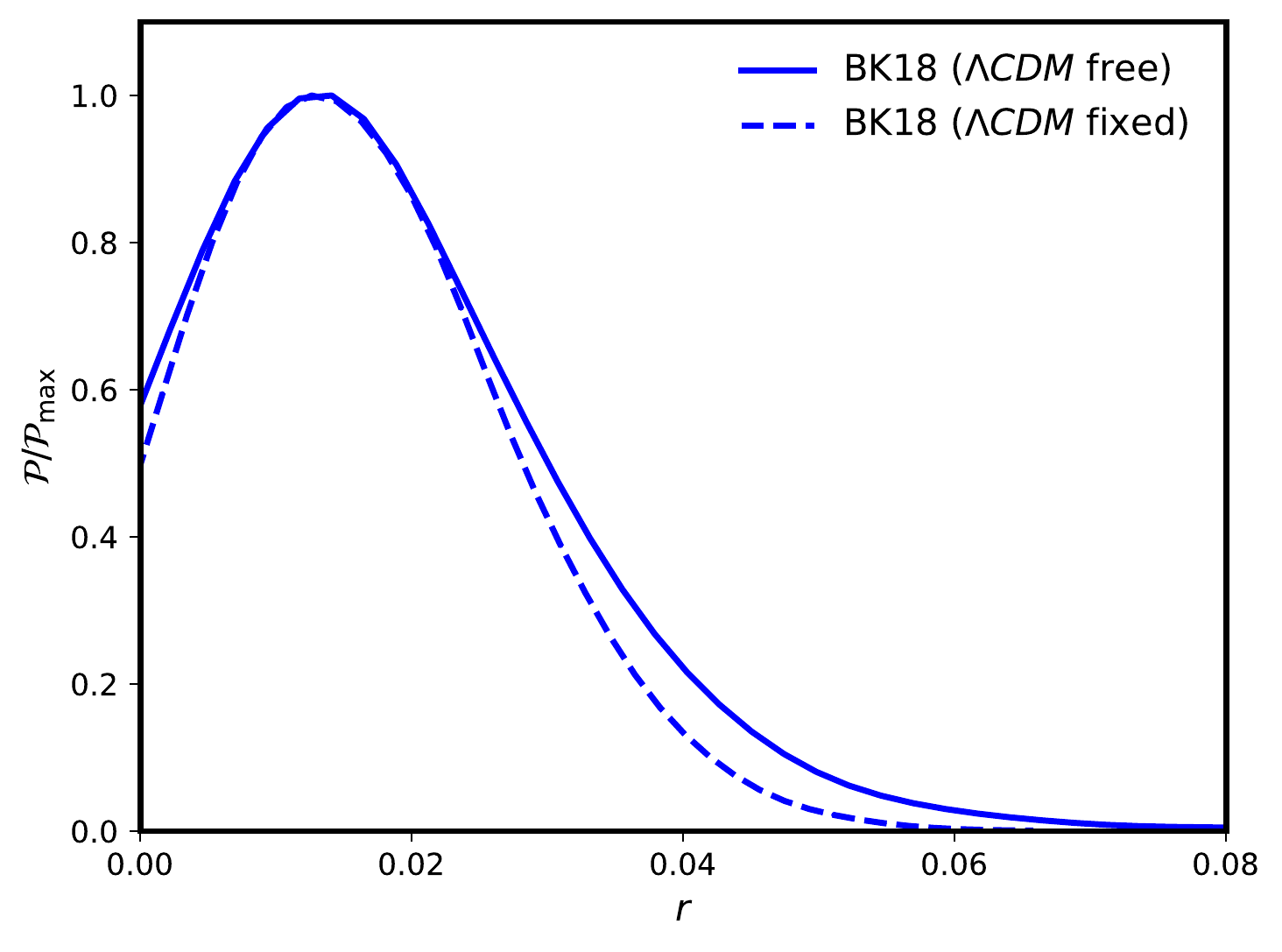}
\caption{Posterior distribution for the tensor-to-scalar ratio $r$, showing the impact of marginalization over \LCDM\ parameters.}
\label{fig:r}
\end{figure}

In the baseline analysis described in Ref.~\cite{BK18}, the BICEP/Keck Collaboration fixed the cosmology to that of best-fitting model from \planck\ 2018 and quote an upper limit of $r<0.036$ at 95\%~CL.
Within this baseline, the uncertainty on \LCDM\ parameters was not propagated, reducing the width of the posterior for the tensor-to-scalar ratio $r$. 
We find that when fitting the BK18 data for \LCDM\ parameters in addition to $r$, the uncertainty on $r$ slightly increases (as illustrated in Fig.~\ref{fig:r}) because the \LCDM\ parameters except for $A_{\rm s}$ are poorly constrained.

The constraints on $r$ then become
\begin{eqnarray}
r = 0.014_{-0.011}^{+0.012} &&\quad \text{(BK18 with \LCDM\ fixed)},\label{eq:BK18fixed}\\
r = 0.015_{-0.013}^{+0.015} &&\quad \text{(BK18 with \LCDM\ free)},
\end{eqnarray}
all compatible with zero\footnote{Uncertainties in Eq.~\ref{eq:BK18fixed} are slightly larger than those in Ref.~\cite{BK18}, despite using the same likelihood. This small difference could be due to assuming different values for the reference \LCDM\ model parameters (we used \planck2018 TT,TE,EE+lowE+lensing~\cite{planck2016-l06}), or might arise from using different MCMC/Boltzmann solver codes.} and resulting in the following upper-limits at 95\,\% CL:
\begin{eqnarray}
r < 0.036 &&\quad \text{(BK18 with \LCDM\ fixed)};\\
r < 0.042 &&\quad \text{(BK18 with \LCDM\ free)}.
\end{eqnarray}
\\

\section{Combining \planck\ and BICEP/Keck} 
The addition of \planck\ data allows us to constrain \LCDM\ parameters, thus reducing the uncertainty on $r$. This was mentioned in a secondary analysis of~Ref.~\cite{BK18} (their appendix~E.1), yielding an upper limit on $r$ similar to that of the baseline when using the earlier \planck\ PR3 data. In this paper, we update the \planck\ data to PR4 and add constraints from the polarized low-$\ell$ likelihood.

\begin{figure}[htbp!]
\centering
\includegraphics[width=8.6cm]{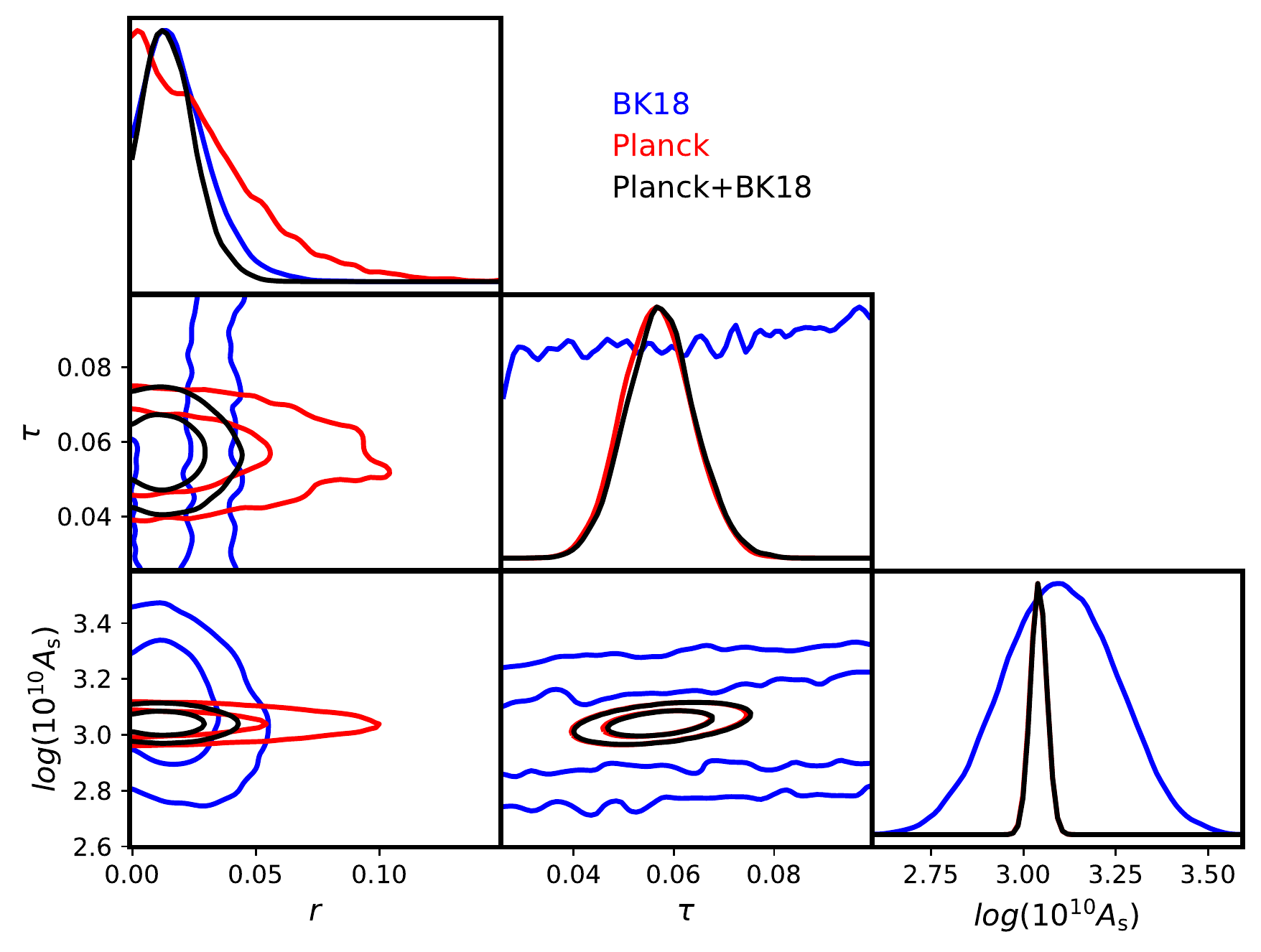}
\caption{Posterior distributions for $\tau$, $A_{\rm s}$, and $r$ using \BK{18}~\cite{BK18}, \planck~\cite{planck_tensor}, and the combination of the two.}
\label{fig:tau-r}
\end{figure}

With the new BICEP/Keck data set included, the uncertainty on $r$ has decreased to $\sigma(r) = 0.014$. We may compare this to the \planck\ uncertainty $\sigma(r) = 0.056$ based on polarized low multipoles; this uncertainty reduces to $\sigma(r)=0.024$ when the $TT+TE+EE$ high multipole data are included as well.
The addition of \lowl\ from \planck\ polarization modes allows the degeneracy with $\tau$ to be broken and also slightly reduces the width of the posterior distribution for $r$. This is illustrated in Fig.~\ref{fig:tau-r}.

The resulting constraint on $r$ using a combination of \planck\ and \BK{18} data tightens to
\begin{eqnarray}
    r = 0.014_{-0.009}^{+0.011} &&\quad \text{(Planck+BK18)},
\end{eqnarray}
which corresponds to $r < 0.034$ at 95\,\% CL. The reionization optical depth is found to be
\begin{eqnarray}
    \tau = 0.057 \pm 0.007.
\end{eqnarray}

\begin{figure}[htbp!]
\centering
\includegraphics[width=0.95\linewidth]{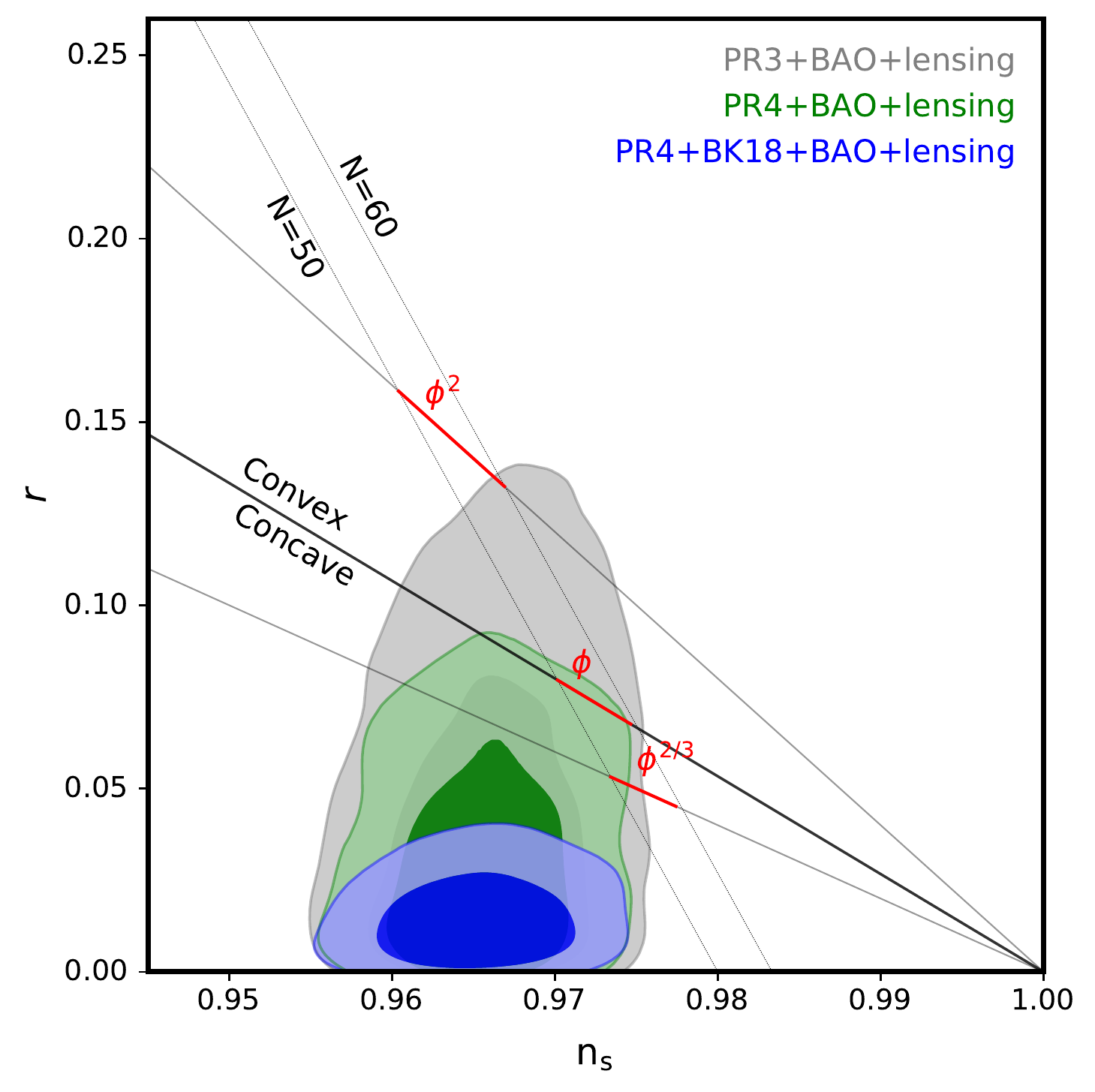}
\caption{
Constraints in the tensor-to-scalar ratio $r$ versus scalar spectral index $n_{\rm s}$ plane for the \LCDM\ model, using CMB data in combination with baryon acoustic oscillation (BAO) and CMB lensing data. The CMB data are \planck\ PR3 (TT,TE,EE+lowE, gray contour), \planck\ PR4 \cite{planck_tensor} (TT,TE,EE+lowlEB, green contour), and \planck\ PR4 joint with BK18 \cite{BK18} (blue contour, this paper).
These constraints assume the inflationary consistency relation and negligible running. Dotted lines show the loci of approximately constant e-folding number $50 < N < 60$, assuming simple $V \propto (\phi/m_{\rm Pl})^p$ single-field inflation. Solid lines show the approximate $n_{\rm s}$--$r$ relation for locally power-law potentials, to first order in slow roll. The solid black line (corresponding to a linear potential) separates concave and convex potentials. This plot is adapted from figure~28 in Ref.~\cite{planck2016-l06}.}
\label{fig:nsr}
\end{figure}

The combination of the two data sets allows us to cover the full range of multipoles that are most sensitive to tensor modes. In combination with baryon acoustic oscillation (BAO \cite{SDSSdr16}) and CMB lensing~\cite{planck2016-l08} data, we obtain an improved upper limit of 
\begin{eqnarray}
    r<0.032 \quad \text{(95\% CL).}
\end{eqnarray}
In the $n_{\rm s}$--$r$ plane (Fig.~\ref{fig:nsr}), the constraints now rule out the expected potentials for single-field inflation (strongly excluding $V\propto\phi^2$, $\phi$, and even $\phi^{2/3}$ at about 5$\,\sigma$).

\section{\label{sec:conclusion}Conclusions}
We have derived constraints on the tensor-to-scalar ratio $r$ using the two most sensitive data sets to date, namely BICEP3 and \planck\ PR4. The BICEP/Keck Collaboration recently released a likelihood derived from their data up to the 2018 observing season, demonstrating a sensitivity on $r$ of $\sigma_r = 0.013$, covering the multipole range from $\ell = 20$ to 300~\cite{BK18}. Complementary \planck\ PR4 data released in 2020~\cite{planck2020-LVII} provide information on the large scales, with a polarized likelihood covering the multipole range from $\ell =2$ to $\ell = 150$~\cite{planck_tensor}. This has poorer sensitivity, with $\sigma_r = 0.024$, but offers independent information, with the constraint on $r$ coming from a combination of $TT$, $TE$, and large-scale $E$ and $B$ data. It is interesting to note that constraints derived purely from temperature anisotropies alone are not competitive anymore ($\sigma_r = 0.1$ \cite{planck_tensor}), since those data are dominated by cosmic variance.

\begin{figure*}[htbp!]
\centering
\includegraphics[width=0.95\linewidth]{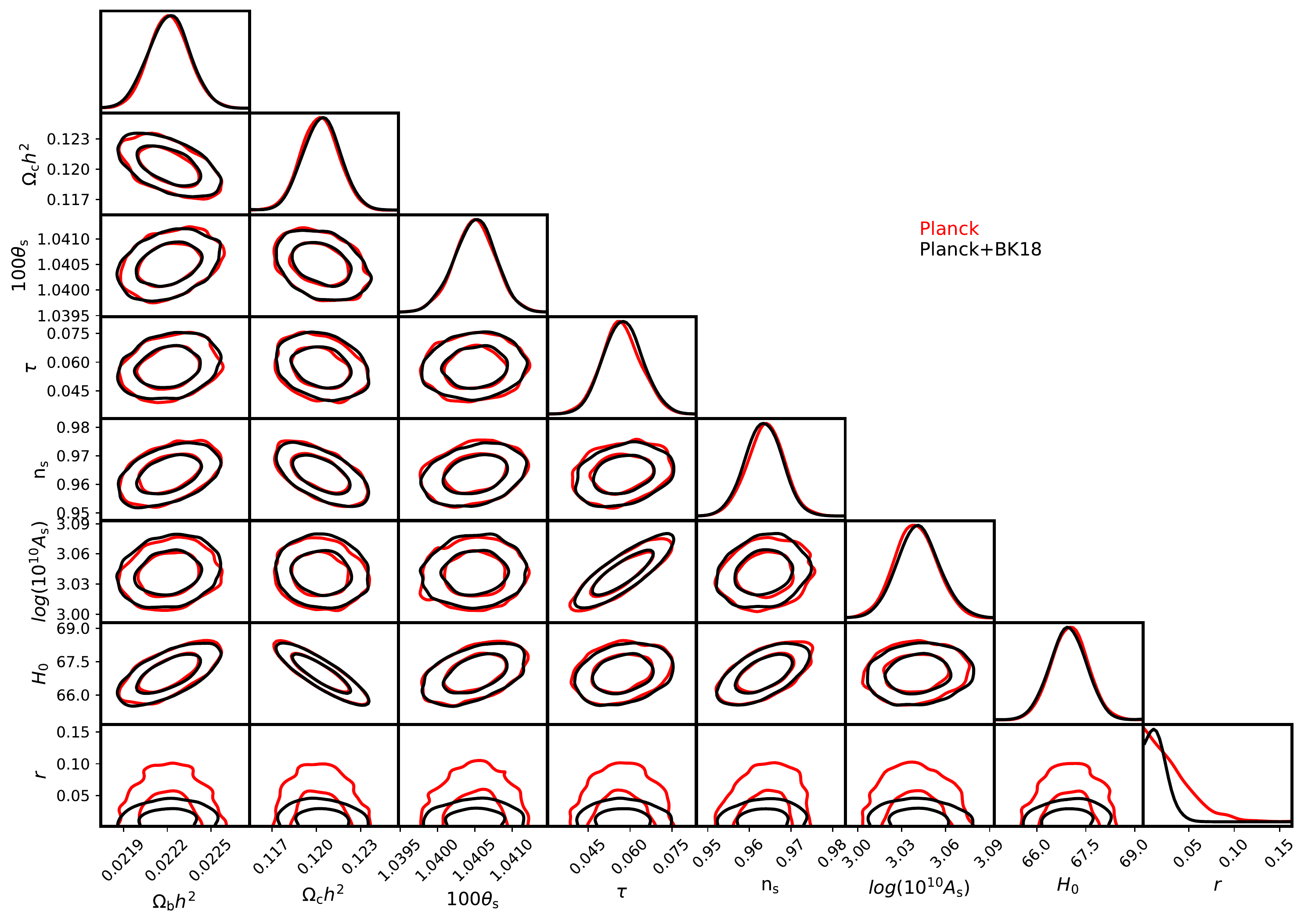}
\caption{Constraint contours (at 68 and 95\,\% confidence) on parameters of a \LCDM+r model using \planck\ (red) and \planck+BK18 (black).}
\label{fig:lcdm}
\end{figure*}

The addition of \planck\ data (including large angular scales in polarization, as well as small angular scales in $TT$ and $TE$) allows us to increase the sensitivity on $r$, as well as to break the degeneracy with the usual six parameters of the \LCDM\ model. 
We find that other \LCDM\ parameters are not affected by the addition of BK18 data (Fig.~\ref{fig:lcdm}).
Combining \planck\ PR4 and \BK{18}, we find an upper limit of $r<0.034$, which tightens to $r<0.032$ when adding BAO and CMB lensing data.

Ground-based experiments (such as BICEP/Keck, the Simons Observatory \cite{SO}, and later CMB-S4 \cite{S4}) will observe the sky with ever deeper sensitivity, placing even stronger constraints on the tensor-to-scalar ratio $r$ (or detecting primordial $B$ modes of course). 
However, improved measurements of the \LCDM\ parameters are essential to achieve strong constraints on $r$. In particular reionization optical depth require very large scales, which are extremely difficult to measure from ground. The next generation of polarized CMB space missions (including LiteBIRD~\cite{litebird}) will be able to deliver $\tau$ with a precision dominated by cosmic variance.

\begin{acknowledgments}
\planck\ is a project of the European Space Agency (ESA) with instruments provided by two scientific consortia funded by ESA member states and led by Principal Investigators from France and Italy, telescope reflectors provided through a collaboration between ESA and a scientific consortium led and funded by Denmark, and additional contributions from NASA (USA).
We gratefully acknowledge support from the CNRS/IN2P3 Computing Center for providing computing and data-processing resources needed for this work.
This research used resources of the National Energy Research Scientific Computing Center, which is supported by the Office of Science of the U.S. Department of Energy under Contract No.\ DE-AC02-05CH11231.
Part of the research was carried out at the Jet Propulsion Laboratory, California Institute of Technology, under a contract with the National Aeronautics and Space Administration (80NM0018D0004).
\end{acknowledgments}

\appendix
\section*{Appendix: Statistical discussions}
The \planck\ likelihood used in this analysis is described in detail in Ref.~\cite{planck_tensor}. It is based on the $N=400$ simulations provided with the \planck\ PR4 data. Those simulations have been shown to be the most realistic description of the \planck\ data, including all relevant systematic effects \cite{planck2020-LVII}.
Using the \planck\ data, we expect correlations at very \lowl, related to long-term systematics, residuals from foregrounds, and cut-sky effects. These should not be neglected.

The covariance used in the likelihood has been constructed from the simulations mentioned above, ensuring the propagation of statistical and systematic uncertainties up to the fitted parameters.
Nevertheless, the limited number of available simulations induces an uncertainty on the estimated covariance of the order of 5\,\% ($1/\sqrt{N}$).
The robustness of the covariance matrix has been checked in two different ways.

\begin{figure}[htbp!]
\centering
\includegraphics[width=0.48\linewidth,height=3.5cm]{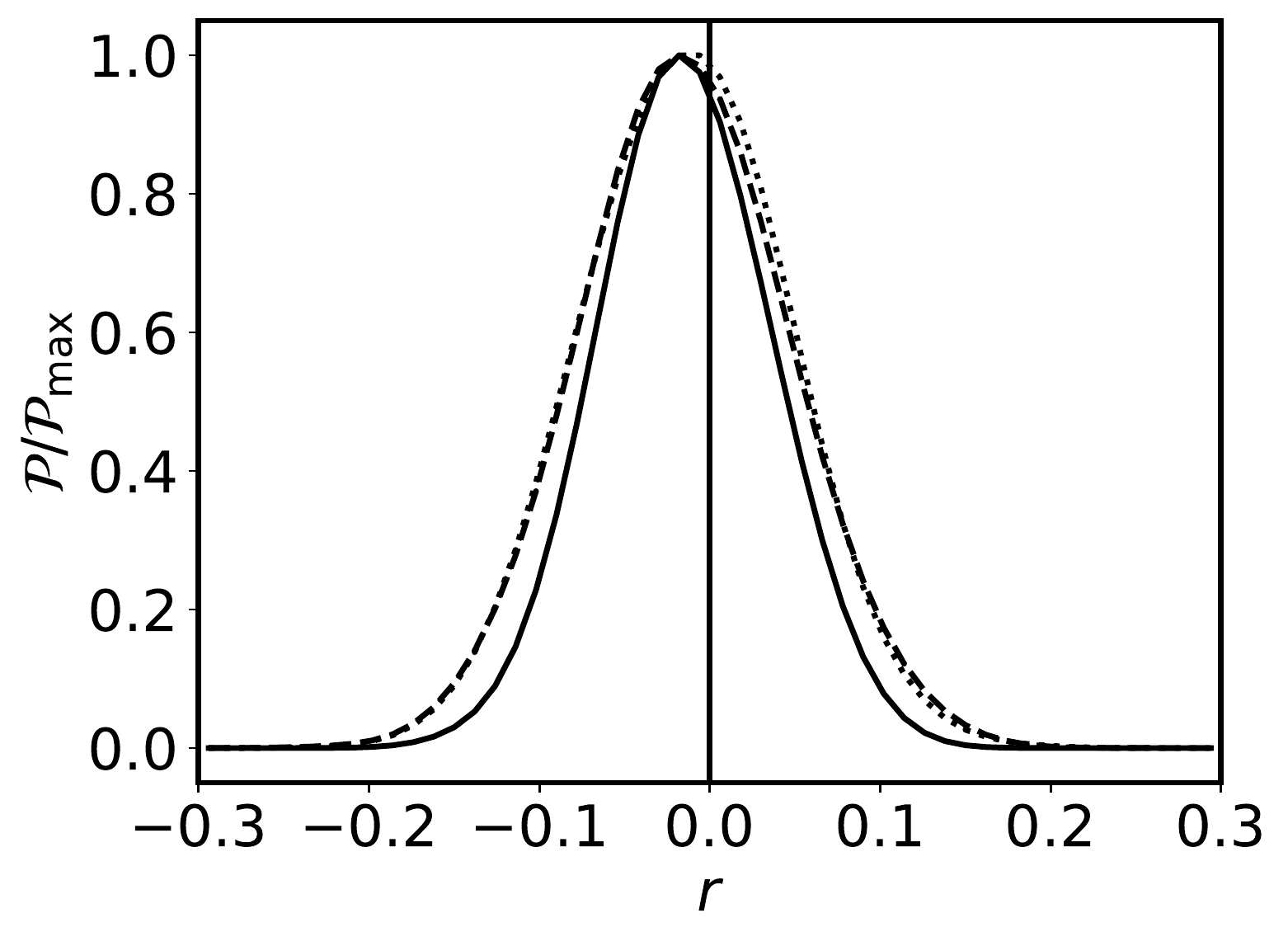}
\includegraphics[width=0.48\linewidth,height=3.5cm]{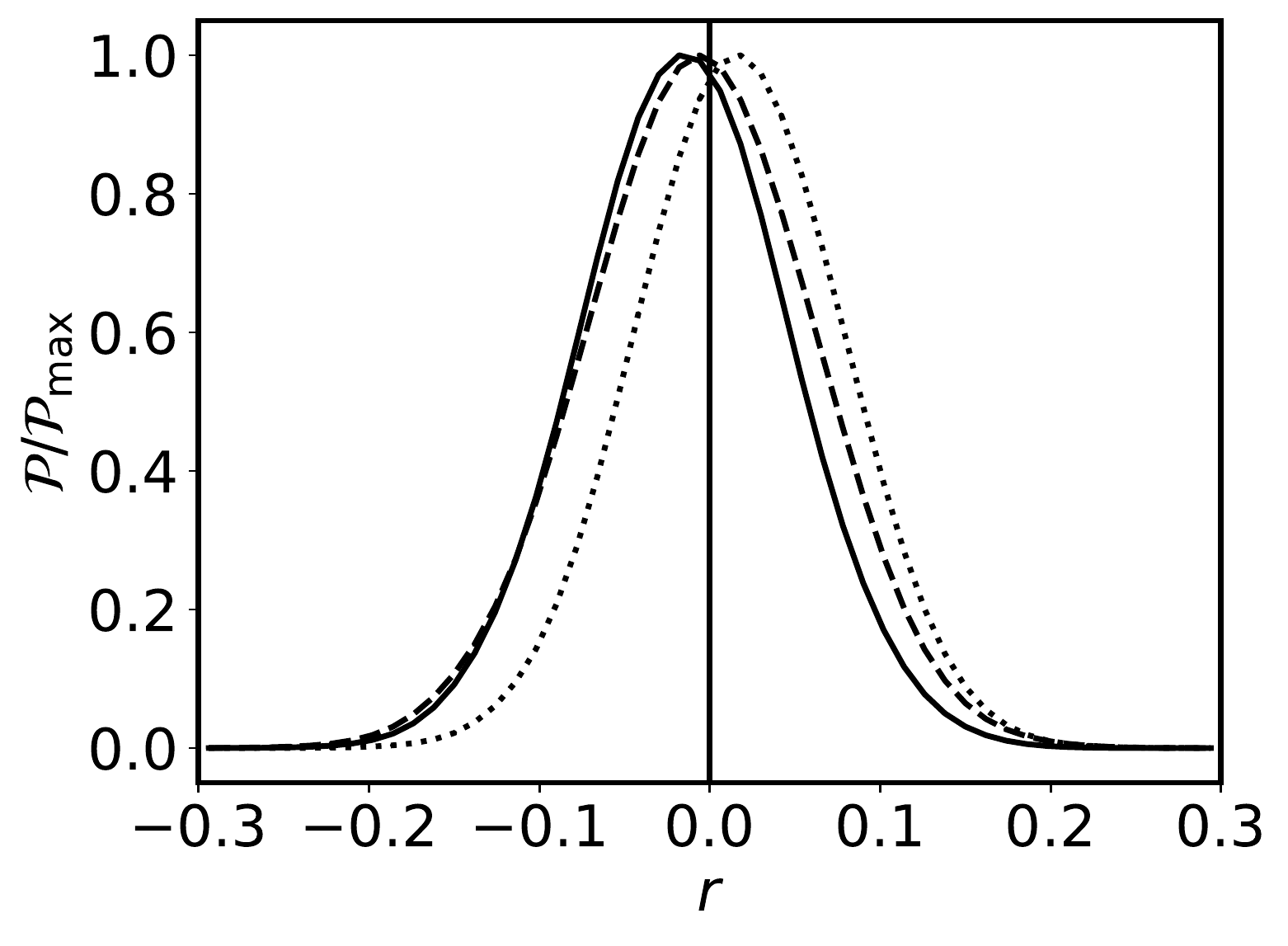}
\caption{Posteriors for $r$. \textit{Left}: after marginalizing over the true covariance matrix (dashed line) and correcting the inverse covariance matrix (dotted line), compared to the effective covariance (solid line). \textit{Right}: using covariance estimates based on the first or last 200 simulations, compared to the effective covariance with 400 simulations (solid line).}
\label{fig:cov}
\end{figure}

Firstly, we marginalized over the unknown true covariance matrix, as described in Ref.~\cite{sellentin16}. The recovered maximum posterior is unchanged, while the width of the posterior is slightly enlarged, as expected due to the marginalization (see left panel of Fig.~\ref{fig:cov}). We also applied the correction on the inverse covariance estimate, as proposed in Refs.~\cite{sellentin16} and \cite{hartlap08}, recovering the same result.

Secondly, we ran the same chains using covariance estimates based on only 200 simulations (right panel of Fig.~\ref{fig:cov}). The posterior distributions of $r$ reconstructed from the \lowlEB\ likelihood using covariance estimates based on either the first or the last 200 simulations are compatible, given the statistical deviations from the covariance matrix estimates.

\begin{figure}[ht!]
\centering
\includegraphics[width=0.9\linewidth]{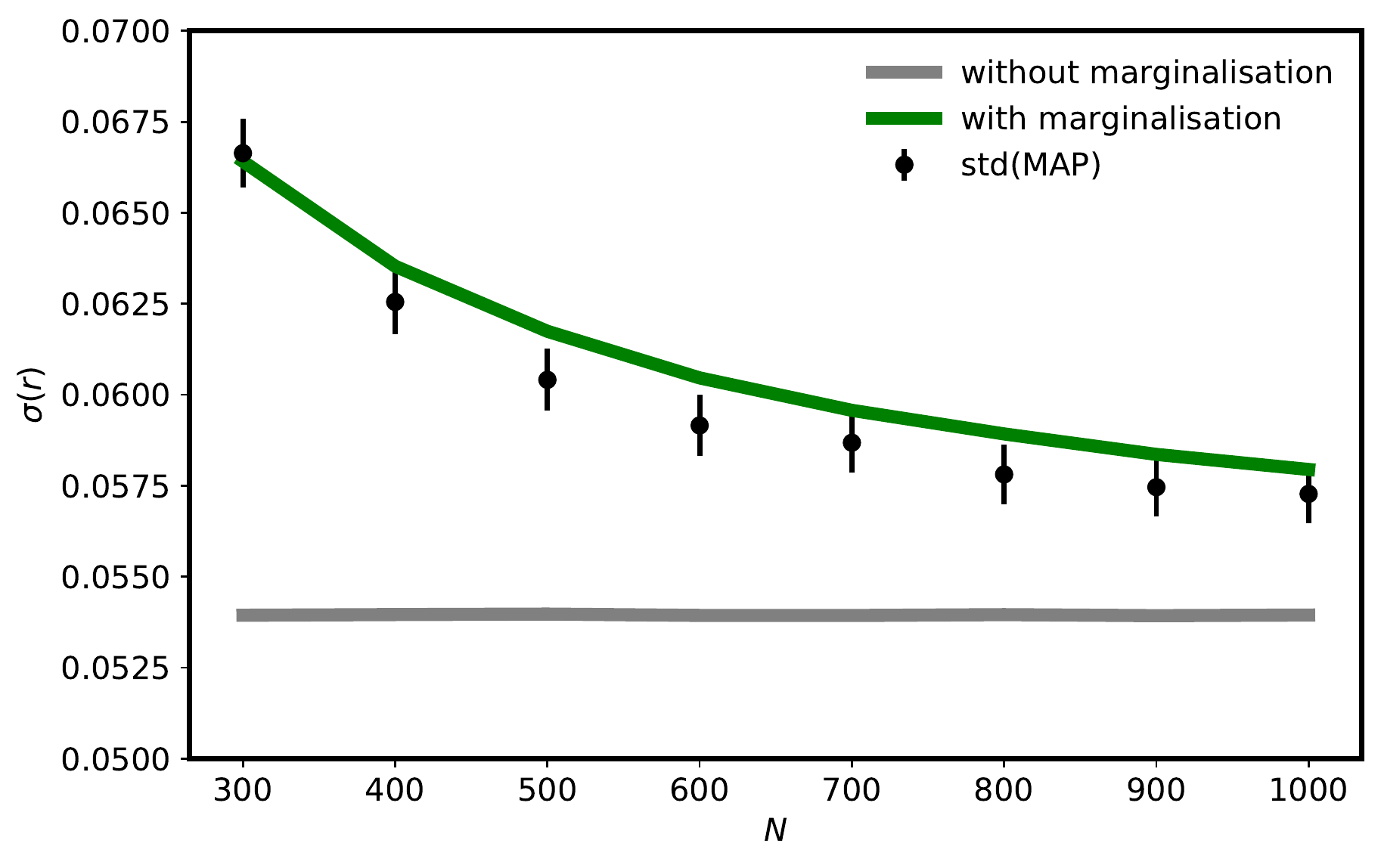}
\caption{$1\,\sigma$ uncertainties estimated without (gray solid line) and with (green solid line) marginalization over the covariance matrix, compared to the standard deviation of the MAP.}
\label{fig:lik_sig}
\end{figure}

We built a Monte Carlo toy model in order to check potential biases in the recovered constraints. We found that the distribution of the maximum a posteriori probability (MAP) peaks at the input value, ensuring that the likelihood is not biased. We found that the marginalization over the unknown covariance matrix ensures that the additional uncertainty coming from the covariance estimation is properly propagated throughout the parameter constraints (see Ref.~\cite{sellentin16}). 
This is illustrated in Fig.~\ref{fig:lik_sig}, where we show that the estimated width of the posterior distribution after marginalization is compatible with the standard deviation of the maximum a posteriori probability. On the contrary, the Hamimeche \& Lewis likelihood without marginalization significantly underestimates the uncertainty for low $N$.

While there may be concern that this could induce a bias in the derived upper limit, we have verified that the recovered upper limits are underestimated by more than 10\,\% compared to upper limits computed with the input covariance in less than 6\,\% of the realisations.

We conclude that covariance matrices based on full end-to-end simulations can be successfully used in likelihoods to infer parameters. The final uncertainty then depends on the number of simulations used to estimate the covariance, but this can be properly taken into account after marginalizing over the true covariance matrix.

\bibliographystyle{apsrev4-2}
\bibliography{tensor2}

\end{document}